\documentclass[showpacs]{revtex4}
\usepackage{amssymb,amsmath}
\usepackage{graphicx}
\usepackage{enumerate}
\usepackage{color}

\begin{document}

\title{Boson stars from a gauge condensate}
\author{Vladimir Dzhunushaliev
\footnote{Senior Associate of the Abdus Salam ICTP}} 
\email{dzhun@krsu.edu.kg} \affiliation{Dept. Phys. and Microel. 
Engineer., Kyrgyz-Russian Slavic University, Bishkek, Kievskaya Str. 
44, 720021, Kyrgyz Republic}

\author{Kairat Myrzakulov}
\affiliation{Institute of Physics and 
Technology, 050032, Almaty, Kazakhstan}

\author{Ratbay Myrzakulov}
\email{cnlpmyra@satsun.sci.kz} \affiliation{Institute of Physics and 
Technology, 050032, Almaty, Kazakhstan}


\begin{abstract}
The boson star filled with two interacting scalar fields is investigated. The scalar fields can be considered as a gauge condensate formed by SU(3) gauge field quantized in a non-perturbative manner. The corresponding solution is regular everywhere, has a finite energy and can be considered as a quantum SU(3) version of the Bartnik - McKinnon particle-like solution.

\end{abstract}

\maketitle

\section{Introduction}

Non-Abelian solitons play an important role in gauge theories of elementary particle physics \cite{Coleman75,Actor79,Rajaraman82}. Therefore it is  reasonable to study  gravitating gauge fields. The first example of gravitating non-Abelian solitons was  discovered by Bartnik and McKinnon in the four-dimensional Einstein-Yang-Mills theory for the gauge group SU(2) \cite{Bartnik88}. Soon after the BK discovery it was  realized that, apart from solitons, the Einstein-Yang-Mills model also contains non-Abelian black holes \cite{Volkov89,Kunzle90,Bizon90}. In Ref. \cite{kuenzle} it was shown that the SU(3) branch of similar solutions exists also. Such solutions can be thought of as eigenstates of  non-linear eigenvalue problems, which  accounts for the discreteness of some their parameters.
\par
The most surprising fact here is the existence of the Bartnik - McKinnon particles, which have not any analogue in flat space. From the physical point of view it proceeds from the fact that 
the classical Yang - Mills equations have not any topologically trivial regular solutions. In fact, it is connected with the confinement problem in quantum chromodynamics. The standard opinion will be that \emph{only} quantized SU(3) gauge field may form regular objects, for example, a hypothesized flux tube connecting two interacting quarks. 
\par 
However, it is known \cite{CS77} that in such sourceless non--Abelian gauge theories 
have no classical glueballs  \cite{R77} which otherwise would be an indication for 
the occurrence of confinement in the quantized theory. The reason simply is that nearby  small portions of the Yang--Mills fields always point at the same direction in internal  space and therefore must repel each other as similar charges. 
\par 
Therefore, it is interesting to investigate gravitating quantum non-Abelian gauge fields. 
The main problem here is to describe quantum non-Abelian gauge fields in a 
non-perturbative manner. In Ref. \cite{Dzhunushaliev:2003sq} is offered an approximate non-perturbative approach to quantization of SU(3) gauge field. It is shown that using 
some assumptions and approximations one can reduce the SU(3) Lagrangian to the Lagrangian describing two interacting scalar fields. 
\par
Thus, in this Letter we will investigate a boson star filled with two interacting scalar fields which can be considered as a gauge condensate. The boson star was first discovered theoretically by Kaup \cite{kaup} and by Ruffini and  Bonazzola \cite{ruffini} (for reviewing the boson star, see Refs.\cite{liddle} and  \cite{mielke}, and other references are cited therein). 
\par 
In Ref. \cite{Brihaye:2004nd} the boson star is considered filled with the complex doublet of scalar fields coupled to an SU(2) non-abelian gauge field. 
\par
In Ref. \cite{Guzman:2005bs} the emission spectrum from a simple accretion disk model around a compact object is compared with the cases of a black hole and a boson star. It was found that, for certain values of the boson star parameters, it is possible to produce spectra similar to those which are generated when the central object is a black hole. 
\par 
One can say that the presented here solution is a quantum SU(3) version of the Bartnik - McKinnon particle-like solution.

\section{Initial equations}

Let us consider Einstein gravity interacting with two scalar fields $\phi, \chi$. The metric is 
\begin{equation}
	ds^2 = \left[ 1 + \frac{M(r)}{r} \right] dt^2 - 
	\displaystyle \frac{e^{\nu(r)}}{1 + \displaystyle \frac{M(r)}{r}} dr^2 - 
	r^2 \left(
		d \theta^2 + \sin^2 \theta d \varphi^2
	\right),
\label{sec2-10}
\end{equation}
The Lagrangian for scalar fields $\phi$ and $\chi$ is 
\begin{equation}
	\mathcal L = \frac{1}{2} \nabla_\mu \phi \nabla^\mu \phi + 
	\frac{1}{2} \nabla_\mu \chi \nabla^\mu \chi - V(\phi, \chi)	,
\label{sec2-20}
\end{equation}
where $\mu, \nu= 0,1,2,3$. The potential $V(\phi, \chi)$ is 
\begin{equation}
	V(\phi, \chi)	= \frac{\lambda_1}{4} \left(
		\phi^2 - m_1^2
	\right)^2 + 
	\frac{\lambda_2}{4} \left(
		\chi^2 - m_2^2
	\right)^2 - \frac{\lambda_2}{4} m_2^4 + \phi^2 \chi^2 ,
\label{sec2-30}
\end{equation}
where $\frac{\lambda_2}{4} m_2^4$ is the constant which can be considered as a  cosmological constant $\Lambda$. In Ref. \cite{Dzhunushaliev:2003} it is shown that these scalar fields present a quantum gauge field. In short it can be shown by the following way. In quantizing strongly interacting SU(3) gauge fields - via Heisenberg's non-perturbative method \cite{heisenberg} one first replaces the classical fields by field operators 
$\mathcal A^B_{\mu} \rightarrow \widehat{\mathcal A}^B_\mu$. This yields the
following differential equations for the operators
\begin{equation}
    \partial_\nu \widehat {\mathcal F}^{B\mu\nu} = 0.
\label{sec2-31}
\end{equation}
where $\mu , \nu = 0,1,2,3$; $B=1,2, \cdots , 8$ are SU(3) color indices. These nonlinear equations for the field operators of the nonlinear quantum fields can be used to determine expectation values for the field operators
$\widehat {\mathcal A}^B_\mu$. One problem in using these equations in order to obtain expectation values like 
$\langle \mathcal A^B_\mu \rangle$, is that these equations involve not only powers or derivatives of $\langle \mathcal A^B_\mu \rangle$ ({\it i.e.} terms like 
$\partial_\alpha \langle \mathcal A^B_\mu \rangle$ or $\partial_\alpha
\partial_\beta \langle \mathcal A^B_\mu \rangle$) 
but also contain terms like 
$\mathcal{G}^{BC}_{\mu\nu} = \langle \mathcal A^B_\mu \mathcal A^C_\nu \rangle$. 
Starting with Eq. \eqref{sec2-31} one can generate an operator differential equation for the product $\widehat {\mathcal A}^B_\mu \widehat {\mathcal A}^C_\nu$ consequently allowing the determination of the Green's function $\mathcal{G}^{BC}_{\mu\nu}$
\begin{equation}
  \left\langle Q \left|
  \widehat {\mathcal A}^B(x) \partial_{y\nu} \widehat {\mathcal F}^{B\mu\nu}(x)
  \right| Q \right\rangle = 0.
\label{sec2-32}
\end{equation}
However this equation will in it's turn contain other, higher order Green's functions. Repeating these steps leads to an infinite set of equations connecting Green's functions of ever increasing order. This construction, leading to an infinite set of coupled, differential equations, does not have an exact, analytical solution and so must be handled using some approximation. The basic approach in this case is to give some physically reasonable scheme for cutting off the infinite set of equations for the Green's functions. Using some assumptions and approximations on 2- and 4-points Green's functions one can reduce the initial SU(3) Lagrangian to an effective Lagrangian describing two interacting scalar fields (for details see Ref. \cite{Dzhunushaliev:2003sq}). 
\par
The scalar fields $\phi$ and $\chi$ which are under discussion here appear in the following way. We assume that in the first approximation two points Green's functions can be calculated as follows 
\begin{eqnarray}
  \left\langle A^a_i (x) A^b_j (y) \right\rangle &=& - \eta_{i j} 
  f^{apm} f^{bpn} \phi^m(x) \phi^n(y),
\label{sec2-33}\\
	\left\langle A^a_0 (x) A^b_0 (y) \right\rangle & \ll & 
	\left\langle A^a_i (x) A^b_j (y) \right\rangle
\label{sec2-34}
\end{eqnarray}
where $A^a_\mu \in SU(2) \subset SU(3), a=1,2,3$; $i,j = 1,2,3$ are spatial indices. And 
\begin{eqnarray}
  \left\langle A^m_i (x) A^n_j (y) \right\rangle &=& - \eta_{i j} 
  f^{mpa} f^{npb} \phi^a(x) \phi^b(y),
\label{sec2-35}\\
	\left\langle A^m_0 (x) A^n_0 (y) \right\rangle & \ll & 
	\left\langle A^m_i (x) A^n_j (y) \right\rangle
\label{sec2-36}
\end{eqnarray}
where $A^m_\mu \in SU(3)/SU(2), m=4,5,6,7,8$. The 4-points Green's functions are a bilinear combination of 2-points Green's functions 
\begin{eqnarray}
  \left\langle A^m_\mu(x) A^n_\nu(y) A^p_\alpha(z) A^q_\beta(u) \right\rangle &=&
  \lambda_1 \biggl[ \left\langle A^m_\mu(x) A^n_\nu(y) \right\rangle 
	  \left\langle A^p_\alpha(z) A^q_\beta(u) \right\rangle + 
  \biggl.,
\nonumber \\
	  &&
 	 \biggl.
	  \frac{\mu_1^2}{4} \left(
	  	\delta^{mn} \eta_{\mu \nu} \left\langle A^p_\alpha(z) A^q_\beta(u) \right\rangle + 
	  	\delta^{pq} \eta_{\alpha \beta} \left\langle A^m_\mu(x) A^n_\nu(y) \right\rangle
	  \right) + 
	  \frac{\mu_1^4}{16} 
	  \delta^{mn} \eta_{\mu \nu} \delta^{pq} \eta_{\alpha \beta} 
	  \biggl] + 
\nonumber \\
	  &&
	  ( \text{permutations of indices} )
\label{sec2-37}
\end{eqnarray}
and 
\begin{eqnarray}
  \left\langle A^a_\mu(x) A^b_\nu(y) A^c_\alpha(z) A^d_\beta(u) \right\rangle &=&
  \lambda_2 \biggl[ \left\langle A^a_\mu(x) A^b_\nu(y) \right\rangle 
	  \left\langle A^c_\alpha(z) A^d_\beta(u) \right\rangle + 
  \biggl.
\nonumber \\
	  &&
	  \biggl.
	  \frac{\mu_2^2}{4} \left(
	  	\delta^{ab} \eta_{\mu \nu} \left\langle A^c_\alpha(z) A^d_\beta(u) \right\rangle + 
	  	\delta^{cd} \eta_{\alpha \beta} \left\langle A^a_\mu(x) A^b_\nu(y) \right\rangle
	  \right) + 
	  \frac{\mu_2^4}{16} \delta^{ab} \eta_{\mu \nu} \delta^{cd} \eta_{\alpha \beta}
	  \biggl] + 
\nonumber \\
	  &&
	 \text{(permutations of indices)}
\label{2f1-30}
\end{eqnarray}
In this Letter we have changed $\phi^a \rightarrow \phi$ and $\chi^m \rightarrow \chi$. 
\par
We consider here the spherically symmetric case and the functions $\phi, \chi$ are $\phi(r), \chi(r)$. The Einstein and scalar field equations are 
\begin{eqnarray}
	R^\mu_\nu - \frac{1}{2} \delta^\mu_\nu R &=& \varkappa T^\mu_\nu ,
\label{sec2-40}\\
	\frac{1}{\sqrt{-g}} \nabla_\mu \left( 
		\sqrt{-g} g^{\mu \nu} \nabla_\nu \phi
	\right) &=& - \frac{\partial V\left( \phi, \chi \right)}{\partial \phi} ,
\label{sec2-50}\\
	\frac{1}{\sqrt{-g}} \nabla_\mu \left( 
		\sqrt{-g} g^{\mu \nu} \nabla_\nu \chi
	\right) &=& - \frac{\partial V\left( \phi, \chi \right)}{\partial \chi} ,
\label{sec2-60}
\end{eqnarray}
where $\varkappa$ is the gravitational constant; $g_{\mu \nu}$ is the metric \eqref{sec2-10} and $g$ is the corresponding determinant. After substituting metric \eqref{sec2-10} for Eq's \eqref{sec2-40} - \eqref{sec2-60} and after algebraic transformations we have the following equations 
\begin{eqnarray}
	\nu' &=& \frac{x}{2} \left(
		\phi'^2 + \chi'^2  
	\right) ,
\label{sec2-70}\\
	M' &=& e^\nu - 1 + \frac{x^2 + M x}{4} \left(
		\phi'^2 + \chi'^2  
	\right) - 
\nonumber \\
	&&
	\frac{x^2}{4}e^\nu  \left[
		\frac{\lambda_1}{2} \left( \phi^2 - m_1^2 \right)^2 + 
		\frac{\lambda_2}{2} \chi^2 \left( \chi^2 - 2 m_2^2 \right) +
		\phi^2 \chi^2
	\right] ,
\label{sec2-80}\\
	\phi'' + \phi' \left( 
		\frac{2}{x} - \frac{\nu'}{2} + \frac{M' - \displaystyle \frac{M}{x}}{x + M}
	\right)&=& \frac{e^\nu}{1 + \displaystyle \frac{M}{x}}\phi \left[
		\chi^2 + \lambda_1 \left( \phi^2 - m_1^2 \right)
	\right] ,
\label{sec2-90}\\
	\chi'' + \chi' \left( 
		\frac{2}{x} - \frac{\nu'}{2} + \frac{M' - \displaystyle \frac{M}{x}}{x + M}
	\right)&=& \frac{e^\nu}{1 + \displaystyle \frac{M}{x}} \chi \left[
		\phi^2 + \lambda_2 \left( \chi^2 - m_2^2 \right)
	\right] ,
\label{sec2-100}
\end{eqnarray}
where $\frac{d (\cdots)}{ dx} = (\cdots)'$; the following functions 
$\phi \sqrt{\varkappa} \rightarrow \phi$, 
$ \chi \sqrt{\varkappa} \rightarrow \chi$, 
constants 
$m_{1,2}\sqrt{\varkappa} \rightarrow m_{1,2}$, 
$\lambda_{1,2}/2 \rightarrow \lambda_{1,2}$ and the dimensionless variable 
$x = r/\sqrt{\varkappa/2}$ are introduced. 
\par 
The boundary conditions are 
\begin{eqnarray}
	\nu(0) &=& 0 ,
\label{sec2-110}\\
	M(0) &=& 0 ,
\label{sec2-120}\\
	\phi(0) &=& \phi_0 , \quad \phi'(0) = 0 , 
\label{sec2-130}\\
	\chi(0) &=& \chi_0 , \quad \chi'(0) = 0 . 
\label{sec2-140}
\end{eqnarray}

\section{Numerical investigation}

\begin{figure}[h]
\begin{minipage}[t]{.45\linewidth}
  \begin{center}
  \fbox{
  \includegraphics[height=5cm,width=7cm]{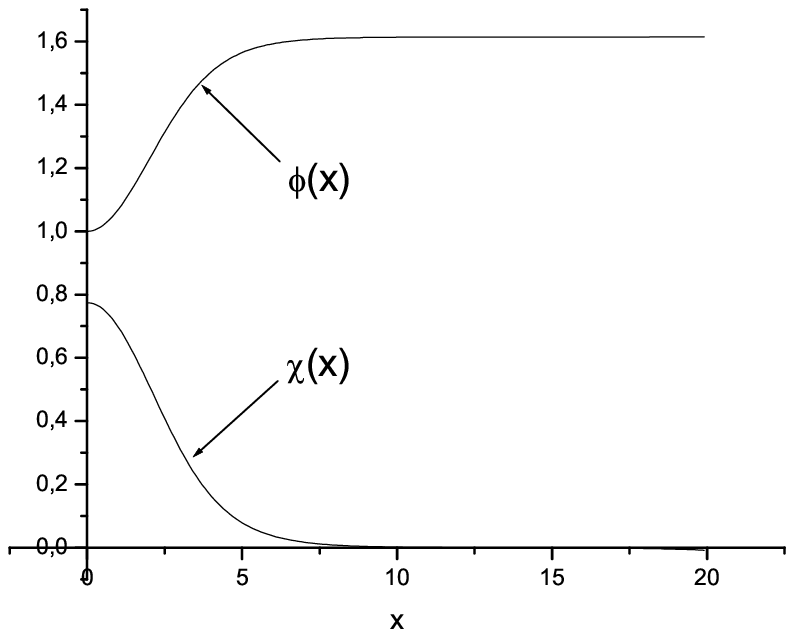}}
  \caption{The functions $\phi^*(x), \chi^*(x)$}
  \label{fig1}
  \end{center}
\end{minipage}\hfill
\begin{minipage}[t]{.45\linewidth}
  \begin{center}
  \fbox{
  \includegraphics[height=5cm,width=7cm]{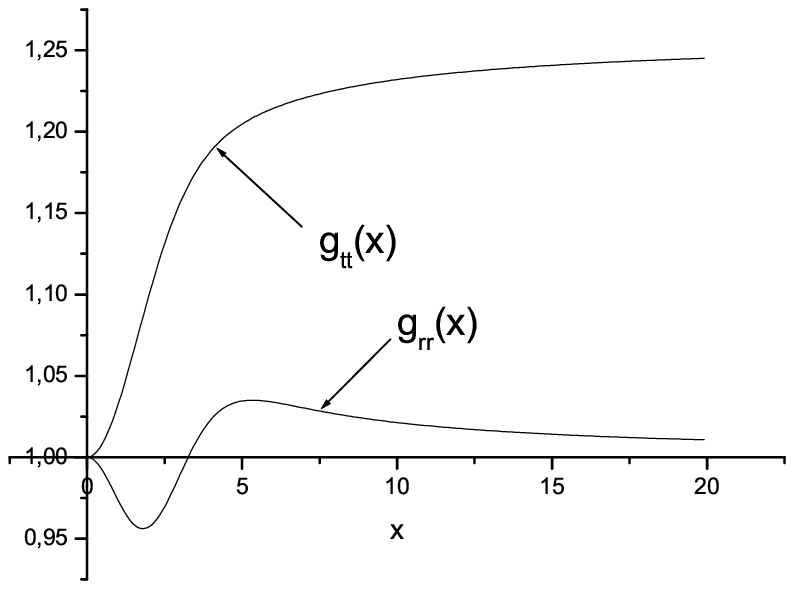}}
  \caption{The profiles of metric components $g_{tt}(x) = 1 + \frac{M(x)}{x}$ and 
  $g_{rr}(x) = \frac{e^{\nu(x)}}{1 + \frac{M(x)}{x}}$.}
  \label{fig2}
  \end{center}
\end{minipage}\hfill 
\end{figure}

For the numerical calculations we choose the following parameters values 
\begin{equation}
	\phi_0 = 1, \quad 
	\chi_0 = \sqrt{0.6}, \quad 
	\lambda_1 = 0.1, \quad 
	\lambda_2 = 1.0 .
\label{sec3-10}
\end{equation}
We apply the methods of step by step approximation to finding of numerical solutions (the details of similar calculations can be found in Ref. \cite{Dzhunushaliev:2003sq}). For the numerical calculations we have to start from a small point $x = \Delta = 0.01$. For this case the boundary conditions \eqref{sec2-110}-\eqref{sec2-140} are 
\begin{eqnarray}
	\nu(\Delta) &=& \nu_4 \frac{\Delta^4}{4} , \quad 
	\nu_4 = \frac{\phi_2^2 + \chi_2^2}{2},
\label{sec2-110}\\
	M(0) &=& M_3 \frac{\Delta^3}{3} , \quad
	M_3 = - \frac{1}{4} \left[
		\frac{\lambda_1}{2} \left( \phi_0^2 - m_1^2 \right)^2 + 
		\frac{\lambda_2}{2} \left( \chi_0^2 - m_2^2 \right)^2 + \phi_0^2 \chi_0^2
	\right],
\label{sec2-120}\\
	\phi(0) &=& \phi_0 + \phi_2 \frac{\Delta^2}{2} , \quad
	\phi'(0) = \phi_2 \Delta , \quad 
	\phi_2 = \frac{\phi_0}{3} \left[
		\chi_0^2 + \left( \phi_0^2 - m_1^2 \right)
	\right],
\label{sec2-130}\\
	\chi(0) &=& \chi_0 + \chi_2 \frac{\Delta^2}{2} , \quad 
	\chi'(0) = \chi_2 \Delta , \quad 
	\chi_2 = \frac{\chi_0}{3} \left[
		\phi_0^2 + \left( \chi_0^2 - m_2^2 \right)
	\right].
\label{sec2-140}
\end{eqnarray}
\begin{enumerate}[{\textbf{Step} }\arabic{enumi}]
	\item On the first step we solve Eq. \eqref{sec2-90} (having zero approximations $\nu_0(x) = 0, M_0(x) = 0$). The regular solution exists for a special value $m^*_{1,i}$ only. For $m_1 < m^*_{1,i}$ the function $\chi_i(y) \rightarrow +\infty$, for $m_1 > m^*_{1,i}$ the function oscillates and $\chi_i(y) \rightarrow 0$
(here the index $i$ is the approximation number). One can say that in this case we solve \emph{a non-linear eiqenvalue problem}: $\chi_i^*(y)$ is the eigenstate and $m_{1,i}^*$ is the eigenvalue on this Step. 
	\item On the second step we solve Eq. \eqref{sec2-100} using zero approximation 
	$\nu_0(x) = 0, M_0(x) = 0$ and the first approximation $\chi_1^*(y)$ for the function $\chi(y)$ from the Step 1. For $m_2 < m^*_{2,1}$ the function $\phi_1(y) \rightarrow +\infty$ and for $m_2 > m^*_{2,1}$ the function $\phi_1(y) \rightarrow -\infty$. Again we have \emph{a non-linear eiqenvalue problem} for the function $\phi_1(y)$ and $m^*_{2,1}$.
	\item On the third step we repeat the first two steps that to have the good convergent sequence $\phi_i^*(y), \chi_i^*(y)$. Practically we have made three approximations. 
	\item On the next step we solve Eq's \eqref{sec2-70} \eqref{sec2-80} which give us the functions $\nu_1(x), M_1(x)$. 
	\item On this step we repeat Steps 1-4 necessary number of times that to have the necessary accuracy of definition of the functions 
$\nu^*(y), M^*(x), \phi^*(x), \chi^*(x)$. 
\end{enumerate}
After Step 5 we have the solution presented in Fig's. \ref{fig1}, \ref{fig2}. These numerical calculations give us the eigenvalues $m_1^* \approx 1.6138771$, 
$m_2^* \approx 1.493441$ and eigenstates $\nu^*(y), M^*(x), \phi^*(x), \chi^*(x)$. 

\par
\begin{figure}[h]
  \begin{center}
   \fbox{
   \includegraphics[height=5cm,width=7cm]{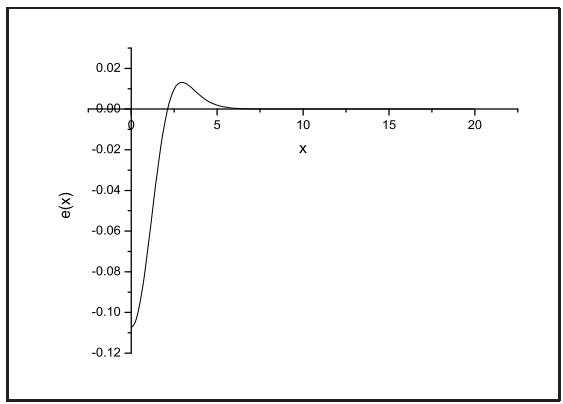}}
  \caption{The profile of energy density $e(x)$.}
  \label{fig3}
  \end{center}
\end{figure}

 
It is easy to see that the asymptotical behavior of the solution is 
\begin{eqnarray}
	\nu(x) &\approx& \max \limits_{x \rightarrow \infty} \left\{
		\lambda_1 m_1^2 \phi_\infty^2 \frac{e^{-2x \sqrt{2 \lambda_1 m_1^2}}}{x},
		\frac{\left( m_1^2 - \lambda_2 m_2^2 \right) \chi_\infty^2}{2} 
		\frac{e^{-2x \sqrt{m_1^2 - \lambda_2 m_2^2}}}{x}
	\right\}, 
\label{sec3-20}\\
	M(x) &\approx& M_\infty x , \quad
	M_\infty = e^{\nu(\infty)} - 1 ,
\label{sec3-25}\\	
	\phi(x) &\approx& m_1 + 
	\phi_\infty \frac{e^{-x \sqrt{2 \lambda_1 m_1^2}}}{x}  ,
\label{sec3-30}\\
	\chi(x) &\approx& \chi_\infty \frac{e^{-x \sqrt{m_1^2 - \lambda_2 m_2^2}}}
	{x} ,
\label{sec3-40}
\end{eqnarray}
where $\nu_\infty, \phi_\infty, \chi_\infty$ are constants. From Fig. \ref{fig2} we see that at the infinity $g_{tt} \rightarrow const \neq 1$. One can avoid this problem in the following way. The numerical calculations show that the asymptotical behavior of the function $M(x)$ is 
\begin{equation}
	M(x) \approx M_\infty x - a 
\label{sec3-41}
\end{equation}
where $a$ is a constant. In this case $g_{tt}$ metric component is 
\begin{equation}
	g_{tt} = 1 + \frac{M}{x} \approx e^{\nu_\infty} - \frac{a}{x}.
\label{sec3-42}
\end{equation}
At the infinity we can rescale the time 
\begin{equation}
	\tilde{t} = e^{-\nu_\infty/2} t 
\label{sec3-43}
\end{equation}
then $g_{tt}$ will be 
\begin{equation}
	\tilde{g}_{tt} \approx 1  - \frac{a e^{-\nu_\infty}}{x}.
\label{sec3-44}
\end{equation}
The dimensionless energy density is 
\begin{eqnarray}
	e(x) &=& T_0^0 = 
	\left( \frac{\varkappa}{2} \right)^2 \varepsilon(r) = 
	\frac{1}{4} \left( 1 + \frac{M(x)}{x} \right) e^{-\nu(x) }
	\Bigl(
		\phi'^2 + \chi'^2 
	\Bigl) + 
\nonumber \\
	&&
		\frac{\lambda_1}{8} \left( \phi^2 - m_1^2 \right)^2 + 
		\frac{\lambda_2}{8} \chi^2 \left( \chi^2 - 2 m_2^2 \right)^2 
		+ \frac{1}{4} \phi^2 \chi^2 
\label{sec3-50}
\end{eqnarray}
and it is presented in Fig. \ref{fig3}. One can calculate the mass of this object in the following manner. The usual definition of the mass $m$ gives us  
\begin{equation}
	\sqrt{\varkappa} m = 4 \pi \int \limits_0^\infty x^2 T^0_0 dx \approx 2.65 
\label{sec3-41}
\end{equation}
here the mass $m$ has the dimension $cm^{-1}$. 

\section{Discussion and conclusions}

In this Letter we have considered two gravitating scalar fields. As the result we have obtained the particle-like solution. The authors point of view (following Ref. \cite{Dzhunushaliev:2003sq}) is that this object is either a star filled with a gravitating quantum SU(3) gauge condensate (if the solution is stable) or a quantum fluctuation in the surrounding gauge condensate (if the solution is unstable). The problem of stability of the presented solution is not simple as the solution is numerical and it is the eigenfunction of non-linear equations system. We hope to investigate this problem in the future research. Nevertheless one can note a remark on this problem. Eq. \eqref{sec3-50} and Fig. \ref{fig3} shows that in the center of the solution there is a negative potential energy density -- the term $- \frac{\lambda_2}{2} m_2^4$ which plays the role of $\Lambda$ term. Such exotic matter can act in a gravitationally repulsive way. In this case the interplay between repulsive forces on small distances and attractive forces on big distances could lead to the stability of the gravitating ball filled with the gauge condensate.
\par
In contrast with the Bartnik - McKinnon solution and non-Abelian black holes the presented solution has a flat space limit which is presented in Ref. \cite{Dzhunushaliev:2003sq}). Another interesting feature is that in both cases the solutions are eigenstates of a nonlinear eigenvalue problem.

\end{document}